\documentstyle[12pt, aasms4, graphics]{article}

\begin{document}

\title{SUPERFLARES ON ORDINARY SOLAR-TYPE STARS}
\author{Bradley E. Schaefer\altaffilmark{1}}
\affil{Department of Physics, Yale University, JWG 463,  New Haven, CT
06520-8121}
\altaffiltext{1}{schaefer@grb2.physics.yale.edu}

\author{Jeremy R. King\altaffilmark{2}}
\affil{Space Telescope Science Institute, 3700 San Martin Drive,
Baltimore, MD 21218}
\altaffiltext{2}{jking@stsci.edu}

\author{\& Constantine P. Deliyannis\altaffilmark{3}}
\affil{Indiana University, Department of Astronomy, 319 Swain West,
Bloomington, IN 47405}
\altaffiltext{3}{con@astro.indiana.edu}

\begin{abstract}

Short duration flares are well known to occur on cool main-sequence stars
as well as on many types of `exotic' stars. Ordinary main-sequence stars
are usually pictured as being static on time scales of millions or
billions of years. Our sun has occasional flares involving up to $\sim
10^{31}$
ergs which produce optical brightenings too small in amplitude to be
detected in disk-integrated brightness. However, we identify nine cases
of superflares involving $10^{33}$ to $10^{38}$ ergs on normal solar-type
stars. That is, these stars are on or near the main-sequence, are of
spectral class from F8 to G8, are single (or in very wide binaries), are
not rapid rotators, and are not exceedingly young in age. This class of
stars includes many those recently discovered to have planets as well as
our own Sun, and the consequences for any life on surrounding planets
could be profound. For the case of the Sun, historical records suggest
that no superflares have occurred in the last two millennia. 

\end{abstract}


\clearpage 
\parskip = 0pt
\marginparsep = 0pt

\section{INTRODUCTION}

        Astronomy has strong imperatives towards looking at exotic sources
and with long exposure times, such that relatively fast variations on
reputedly steady sources can easily be overlooked.  Astronomers have a
poor track record of realizing the existence of new classes of flares
until many years after the first definitive observations, such as with
ordinary flare stars, supernovae, W UMa stars, gamma-ray bursts, and Mira
flares.

        The search for fast flares on stars has two severe problems.
First, a wide variety of instrumental or environmental artifacts can mimic
a brightness increase (Schaefer 1989).  Second, sufficiently rare events
are difficult to detected.  Nevertheless, stellar flares are known to come
from many types of stars ranging from faint red dwarfs to many types of
`exotic' stars (Schaefer 1989; Haisch, Strong, \& Rodon\'{o} 1991).

        We have been collecting reports in the literature on fast flares
as a study of the background for searches of optical flashes from
gamma-ray bursts.  Some of the reported flashes come from G stars similar
to our own Sun.  The next section and Table 1 will describe the flares
observed on nine ordinary main sequence stars within half a spectral class
of our Sun (F8 to G8):

\section{NINE SUPERFLARES}

Groombridge 1830 (HR4550) has the third largest proper motion on the sky
and is the nearest belonging to the old ($> 10$ Gyr) halo population of
our galaxy. It has had extensive series of multiple exposure
photographs taken for astrometric studies. On 27 April 1939, a
six-exposure plate from the Allegheny Observatory showed Groombridge 1830
to be significantly bright on four of its images (Beardsley, Gatewood, \&
Kamper 1974). The brightest image was 0.62 mag brighter than normal (in 
the photographic magnitude system) and the flare duration was slightly 
longer than 18 minutes. Microdensitometry show no change in the
point-spread-function for the flare images, the flare image centers were
within 0.03'' space of the expected position, and all other stars on the
plate were constant. The total flare energy (in the blue band alone) is 
$\sim 10^{35}$ ergs with an uncertainty of a factor of a few due to having
only four points on the light curve. 

$\kappa$ Ceti generally has the 5875.6 $\AA$ He I D3 line in absorption
(Danks \& Lambert 1985), yet it was seen in emission (Robinson \& Bopp
1988) on 24 January 1986 in a spectrum taken at the Ritter Observatory
with an intensified Reticon detector at a resolution of 0.25 $\AA$ and a
signal-to-noise ratio of $> 75$. The flare spectrum had a 40-minute
integration, while another spectrum started 29 minutes later showed no He
I emission. The emission line has all the hallmarks of a real feature;
with a narrow superposed telluric absorption line, the correct central 
wavelength for the radial velocity of $\kappa$ Ceti, the expected shape
for an emission line, no other anomalies in the flare spectrum, and no
similar anomalies on any other spectrum. Big flares on our Sun also show
the He I line in emission (Jefferies, Smith, \& Smith 1959), with no other
spectral changes within the range of the $\kappa$ Ceti spectrum, so the
transient line implies that $\kappa$ Ceti was caught with a big flare.
However, the He I lines appears in emission in our Sun only when the small
flare region is recorded on time scales comparable to the flare duration,
typically 200 seconds (Jefferies, Smith, \& Smith 1959). The $\kappa$ Ceti
spectrum records the {\em whole disk} of the star with an integration 12
times longer than the expected flare duration, so the size of the flare
must have been very large. We can estimate its relative size by comparing 
the equivalent widths of the helium line above background for the $\kappa$
Ceti flare (0.13 $\AA$; Robinson \& Bopp 1988) and a solar flare of
importance 2 (0.72 $\AA$; Jefferies, Smith, \& Smith 1959). However, the
equivalent width of the $\kappa$ Ceti flare must be corrected for the
short duration during the long exposure (a factor of 12) while the
equivalent width of the solar flare must be corrected to that of a whole
disk spectrum (a factor of $\sim 3 \times 10^{-4}$ for the typical area of
an importance 2 flare; see Allen 1976). The corrected ratios of equivalent
widths is then $\sim 7000$, which presumably is comparable to the ratio of
total energies. With a solar flare of importance 2 having $10^{30.4}$ ergs
(Allen 1976), we estimate the flare on $\kappa$ Ceti to involve $\sim 
2 \times 10^{34}$ ergs. 

MT Tau has only one known flare, which was discovered during a search for
flare stars in the Pleiades at Tonantzintla Observatory (Haro \& Chavira
1969). The flare appears as brightenings on a photographic plate through 
a U filter consisting of six 10-minute exposures with slight positional 
offsets between each exposure. G. Haro has rechecked the discovery plate 
and states that there is no doubt that MT Tau underwent a flare (Weaver \& 
Naftilan 1973). MT Tau is not a Pleiad since it has non-member proper
motion, does not show Ca II or hydrogen emission, and is 5 mag too faint
for the Pleiades distance given its spectral classification (Weaver \&
Naftilan 1973). This classification is G5 V (confirmed by W. P. Bidelman),
as measured with the Kitt Peak 84-inch telescope with a dispersion of 200
$\AA \cdot mm^{-1}$(Weaver \& Naftilan 1973). Based on a spectroscopic 
parallax, we adopt a distance of 2200 pc, the flare energy (in the U
band alone) is of order $10^{35}$ ergs with a factor of 10 uncertainty
due to
the limited light curve. 

$\pi^{1}$ UMa was seen to flare in the x-ray band with the imaging
detectors on the EXOSAT satellite (Landini et al. 1986). During a three
hour observation, the star rose from its normal x-ray  brightness to a
peak in less than 8 minutes followed by a decay with an e-folding time
scale of 1000s. The flare was detected independently with multiple 
detectors on EXOSAT at highly significant levels, from an imaged region
centered on the quiescent emission, while the background was stable; thus
giving strong evidence that the star indeed had a flare. Spectral fits are
consistent with an isothermal source of constant emission measure whose
temperature peaked at around $10^{8.0}$ degrees cooling to $10^{7.4}$
degrees. The total energy from 0.1-10 KeV is $2 \times 10 ^{33}$ ergs.

S Fornacis is identified as a variable star for only one incident
(Ashbrook 1959). On 2 March 1899, Lewis Swift had discovered a comet in
the evening twilight, with confirmation from Lick Observatory. European 
observers were notified by telegraph, so as darkness fell on 6 March 1899, 
four observers tried to measure the position of Comet 1899a with respect
to nearby bright stars. Three of the observers all independently reported
one particular comparison star to be roughly 3 magnitudes brighter than
normal, making it hard to recognize the field although the measured
astrometric positions were correct for the quiescent star. These
observers (in Vienna Austria, Arcetri Italy, and Bamberg Germany) all
reported the star bright from 7:37 to 7:54 UT. The observers (J.
Holetschek, A. Abetti, and E. Hartwig) were highly experienced and well 
respected. While the photometric accuracy of the visual reports is not
high, the existence of a 3 magnitude anomaly is a very significant claim. 
The {\em independent} discovery by three widely-separated and skilled 
observers and the three astrometric positions removes all doubt about
that S For was flaring. The duration of the flare is constrained by the
lack of a reported anomaly by E. Millosevich in Rome (who also used 
S For as a comparison star) at 7:13 UT and the normal brightness shown on 
Harvard plate I22535 at 13:20 UT. So the duration of the flare is from 
17 to 367 minutes. The Hipparcos distance to S For is 147 pc with a 25
\% uncertainty, and the flare energy (in the V band alone) is roughly 
$2 \times 10^{38}$ ergs with an uncertainty of about one order of
magnitude due to poor light curve information. 

BD +$10^{\circ}2783$ was fortuitously in the field-of-view of Markarian
841 during a series of observations (George \& Drake 1998) with the ROSAT
PSPC on 21 January 1992. This series consisted of four intervals (with
510, 580, 830, and 830 seconds of data) over a ten hour period. In the
first interval, BD +$10^{\circ}2783$ was brightening from 1.5 to 1.8
counts per second. The second, third, and fourth intervals (starting 
93, 191, and 583 minutes after the first) shows the flux at 0.29, 0.20,
and 0.13 counts per second respectively. The excess counts above
quiescence totaled 1500 in the four observed time intervals. A spectral 
analysis is consistent with a coronal plasma with a dominant temperature
of 1.4 keV. For a distance of 150 pc (based on a spectroscopic parallax),
this corresponds to a flare energy 
from 0.1-2.0 keV of $3.0 \times 10^{34}$ erg for just the four observed 
time intervals alone (George \& Drake 1998). If the time coverage had been 
complete and the flare light curve had an exponential decay, then the
actual x-ray energy would have been an order of magnitude higher.

o Aquilae was monitored with accurate B and V photometry from May 1979 to
September 1980 as part of a broad program following many stars at the
David Dunlap Observatory (Bakos 1983). On two occasions, the star showed 
significant flares. The first flare (20 September 1979) had a V amplitude
of 0.09 mag, no significant change in color, and a duration of $< 5$ days.
The star was measured to be bright in both the B and V bands on a single 
night. The second flare (24 July 1980) had a V magnitude of 0.09 mag, 
no significant color change, and a duration of roughly 15 days. The star
was measured to be bright on four separate nights in both the B and V
filters, although the duration and number of the flares are not well
constrained. The energy of the second flare (in the B and V bands alone)
is estimated to be $9 \times 10^{37}$ erg.

5 Serpentis was also followed photometrically during 1979 and 1980,
showing significant flares on three occasions (Bakos 1983). The V-band
amplitudes were 0.05, 0.09, and 0.07 mag for durations of $<11$, $<25$,
and 3-15 days respectively. The colors for each flare were unchanged from
normal except that the last flare is bluer (with a B-band amplitude of 
0.12). The first two flares were detected on only one night each in both B
and V, while the third flare was detected on two nights also in B and V.
For a three day duration, the flare energy (in the B and V bands alone)
is $7 \times 10^{37}$ erg.

UU CrB was observed as a comparison star for a nearby eclipsing binary
(Olson 1980). The photoelectric measurements were made with five filters
(from ultraviolet to the far red) over 12 nights (16.4 hours total) with 
repeated cycling between variable, comparison star, check star, and sky.
On 21 May 1980, a flare was recorded near the start of its rise, with the
peak 45 minutes later. This peak was followed by a decay with a speed
comparable to that of the rise, although the photometry stopped a dozen
minutes after the peak. A total of 27 independent measures in five
filters (interspersed with check star measures) show the flare. Pairs
of magnitudes in all five filters taken two hours later showed the star
at its quiescent light level. The amplitudes in the u, v, b, y, and
$I_{K}$ filters are 0.18, 0.11, 0.10, 0.05 and 0.30 mag respectively. 
The colors of the excess light are extremely blue shortward of $5500 \AA$
and extremely red longward of $5500 \AA$. The total energy (across the 
observed optical bands) was $7 \times 10^{35}$ ergs (Olson 1980).

\section{SUPERFLARE PROPERTIES}

 The flare energy from our nine stars can be compared to flares on
our Sun. (Distances used for these calculated energies are from Hipparcos
and are accurate to better than 10 \% except as noted in the previous
section.) 
Typical solar flares have energies of $10^{29}$ ergs, the
largest white light flare has roughly $3 \times 10^{31}$ ergs in the
visible, and the largest x-ray flare had an energy of $\sim 2.5
\times 10^{31}$ ergs (Haisch, Strong, \& Rodono 1991). The six optical
flares from Table 1 have energies ranging from $1 \times 10^{35}$ to
$9 \times 10^{37}$ ergs over various fractions of the visible range;
and this is $> 3000 \times$ the most energetic solar flares integrating
over the entire visible range.  Bolometric corrections will substantially
increase the total energy emitted by the flares.  The two x-ray flares
from Table 1 have x-ray energies roughly $100 \times$ and $>> 1000\times$ 
the largest solar x-ray flare.  The spectroscopic flare on $\kappa$ Cet
has an estimated energy $7000 \times$ that of a solar flare of importance
2. In all cases the conclusion is that they are $\geq 100 \times$ that of
the largest solar
event even allowing for the uncertainties in the measured energies.  This
is our justification for calling these flares as `superflares'.

        Could the superflares arise from some previously known stellar
flare mechanism on some unknown star hidden in the system?  Low luminosity
stellar flare sources (ordinary red dwarf flare stars) can easily reside
in the systems without notice, but the observed flare properties are all
wrong for this explanation.  In particular, three of the flares are red or
neutral in color, the two x-ray flares have unreasonable energetics, the
six optical events require implausible amplitudes, while all events have
durations greatly longer than those of flare stars.  Known classes of
stellar flares that are similar to the observed superflares (RS CVn and BY
Dra flares) could arise on companion stars to the primary as part of a
double or triple system, however their luminosity is roughly equal to that
of the primary star and would definitely be identified in the study
reported below.  So the superflares must arise either from a previously
unknown mechanism upon hypothetical companions or arise upon the primary
star.

The nine superflare stars appear to be ordinary stars like our
Sun, but is it possible that they are very young, rapidly rotating, or in
a close binary system?  This is an important question since there are
known classes of stellar flares in systems with G star components that are
characterized by these properties (T Tauri stars, RS CVn stars, BY Dra
stars, and cataclysmic variables).  To answer this question, we have
collected measurements of quantities that are sensitive to youth,
rotation, and duplicity (Table 2).  The spectral type, luminosity class,
the lithium equivalent widths, rotational $v \cdot sin(i)$, and duplicity
are from recent literature referenced in SIMBAD.  We have observed the
lithium equivalent widths and $v \cdot sin(i)$ for S For, UU CrB, and
BD+$10^{\circ}2783$ with the HYDRA spectrograph on the WIYN 3.5 m
telescope in April
1996.  The calcium H \& K activity index (S) is defined in Baliunas
(1995). The distances are all from Hipparcos parallaxes except for the
distances to MT Tau and BD+$10^{\circ}2783$ which are based on
spectroscopic parallaxes.  The x-ray luminosities are from ROSAT for
0.1-2.5 keV. Ordinary stars (as distinct from exotic sources like pre-main
sequence stars, RS CVn stars, BY Dra stars, and cataclysmic variables) can
be reliably distinguished by the simultaneous criteria that the lithium
equivalent width must be $< 100$ $m\AA$, the $v \cdot sin(i)$ is $<10$ $km
\cdot s^{-1}$, the S value is $< 0.5$, the x-ray luminosity is $<3 \times
10^{29} erg \cdot s^{-1}$, and no radial velocity variations are
detectable. All  nine of our superflare stars pass these tests as not
being very young, rapidly rotating, or in a close binary; so we can
definitely rule out all known classes of stellar flares.

        Since these stars are F8-G8, on the main sequence, single (or at
least only in very wide doubles), and have no apparent large differences
from our Sun, we feel justified in calling the superflare stars as
solar-type.  This is not to say that the stars have identical properties
to our Sun, for example Gmb 1830 and 5 Ser are substantially older than
our Sun, while $\kappa$ Cet and $\pi^{1}$ UMa are $\sim 1$ Gyr in age.
Hipparcos parallaxes indicate that three stars (S For, 5 Ser, and o
Aql) might be at or perhaps just barely past the main sequence turnoff.
Nevertheless, two of our stars ($\kappa$ Cet and $\pi^{1}$ UMa) have both
been independently identified by three groups as among the ``best true
solar twins'' and ``solar analogs'' (de Strobel 1996, Gaidos 1998, DeWarf
et al. 1998).

How frequently does an average solar-type star suffer a superflare? With
no adequate systematic studies, any rate estimate must be crude. X-ray 
satellites have logged $\sim 20$ years of imaged observations over
typical fields-of-view of $\sim 1$ square degree with $\sim 1$ G stars
per square degree brighter than $V = 10$ mag (Allen 1976) resulting in 2
reported flares; suggesting a crude average recurrence time scale of a
decade. Of the 9110 naked eye stars in the Yale Bright Star catalog
(Hoffleit 1982), 4\% are main sequence stars with spectral class from F8
to G8 
and five have reported superflares in Table 1. An order of magnitude
estimate is that these stars have been monitored over the last century for
$\sim 1$ year of time for which a superflare could have been detected. If
so, then the average recurrence time scale is of order a century. The
usual long exposure patrol plates are insensitive to flares (Schaefer
1989), but series of short exposures with positional offsets are good
for discovering flashes from stars. Of order $10^{4}$ of these chain 
photographs have been taken, each typically an hour in duration covering 
$\sim 10$ square degrees to typically 15 mag recording $\sim 10^{3}$
solar-type stars (Allen 1976). With these estimates, the two events from
Table 1 detected with chain photography imply a recurrence time scale of 
order 600 years. (The search for gravitational lensing events from MACHOs
has not yet placed any useful limits on the superflare rate because G type
main sequence stars are below the thresholds for bulge and LMC stars and
also because the usual detection criteria requires $\geq 6$ times when the
star is $> 0.32$ mag bright.) While these time scales are accurate only to
within one or two factors of ten, it is clear that we are dealing with an
{\em average} recurrence time of decades or centuries. If only some
fraction of the solar-type stars flare, then the rate might be higher.

What is the frequency of the superflares on our Sun? This need not equal
the {\em average} value for similar stars. Any superflare in the last 150
years of scientific monitoring of the Sun would definitely have been
recognized. Within historical times, a superflare would presumably 
have been recorded as a bright (oddly colored?) Sun or a short intense
heat wave. Perhaps the most general and reportable phenomenon would likely 
be a world-spanning aurora visible to equatorial latitudes. In the absence
of such reports, the Sun has likely not had any superflares in the last
two millennia. On longer time scales, there are few ways to identify all 
but the most powerful superflares, although we are reminded of the Greek
myth of Pha\"{e}thon wherein the chariot of the Sun drove too close to the
Earth and scorched the Sahara desert. In all, we conclude that our Sun has
a significantly lower superflare frequency than the average for the stars
in our sample, and perhaps the Sun's rate is zero.

\section{CONSEQUENCES}

Both superflares and planets are common around normal solar-type
stars, so it is natural to examine the consequences of superflares on
planets.  For a superflare with a one hour characteristic duration, the
luminosity from the flare will equal the normal luminosity from the star
for a $10^{37}$ erg event.  The flux deposited by a single flare on a
planet is $3.5 \times 10^{7} erg \cdot cm^{-2}$ for a $10^{35}$ erg
superflare at a distance of 1 AU. On a rocky surface, this flux is greatly
too low to cause melting or other geophysical effects.  On an icy surface,
this flux will lead to large scale melting and the formation of flood
plains for superflares with greater than $ \sim 10^{38}$ erg (assuming a
composition dominated by water ices, a characteristic absorption depth of
1 cm, and an orbit with radius 5 AU).

        The effects of a superflare on a planet with an atmosphere will
depend critically on the energy and spectrum of the flare and on the
structure of the atmosphere.  Possible effects include temporary heating,
worldwide aurora, the temporary break up of an ionosphere, and ozone
depletion.  The ionizing radiation of gamma rays, x-rays, and energetic
protons will not reach the ground but will be absorbed in the upper
atmosphere.  For an Earth-like atmosphere, this will create nitrous oxides
at high altitude which will start a cycle of ozone destruction that will
last long past the end of the superflare.  An event with $\sim 10^{36}$
ergs of ionizing energy will result in roughly 80\% ozone loss for longer
than a year, with normal stellar ultraviolet light then irradiating the
surface (Ruderman 1974).

        With either a causal (Rubenstein \& Schaefer 1999) or a casual
connection between planets and superflares, the superflares might play a
significant role in the evolution of any life on the surface of planets or
their moons.  From the last paragraph, the effects of temperature rises
and ultraviolet light at the surface could prove to be damaging to
existing life, perhaps to the extent of causing extinctions.
Alternatively, the superflares might provide an energy source to create
organic molecules (like the lightning in the Miller-Urey experiment) over
the star-facing hemisphere as the first step in the creation of life.

\begin{table}
\caption{{\bf Superflares.}}
\begin{center}
\begin{tabular}{llllll}
\hline
Star & Detector & $V_{normal}$ & Amplitude & Duration & Energy(erg)\\
\hline
Gmb 1830 & Photography & 6.45 & $\Delta B = 0.62 mag$ & 18 min & $E_{B}
\sim 1 \times 10^{35}$ \\
$\kappa$ Cet & Spectroscopy & 4.83 & $EW(He) = 0.13 \AA$ & $\sim 40$
min & $E \sim 2 \times 10^{34}$ \\
MT Tau & Photography & 16.8 & $\Delta U = 0.7 mag$ & $\sim 10$ min &
$E_{U} \sim 1 \times 10^{35}$ \\
$\pi^{1}$ UMa & X-ray & 5.64 & $L_{x} = 10^{29} erg/s$ & $>\sim 35$ min &
$E_{X} = 2 \times 10^{33}$ \\
S For & Visual & 8.64 & $\Delta V \sim 3$ mag & 17-367 min & $E_{V} \sim
2 \times 10^{38}$\\
BD $+10^{\circ}2783$ & X-ray & 10.0 & $L_{X} = 2 \times 10^{31} erg/s$ &
$\sim 49$ min & $E_{X} \gg 3 \times 10^{34}$ \\
o Aql & Photometry & 5.11 & $\Delta V = 0.09 mag$ & $\sim 5 - 15$ days &
$E_{BV} \approx 9 \times 10^{37}$ \\
5 Ser & Photometry & 5.06 & $\Delta V = 0.09 mag$ & $\sim 3 - 25$ days &
$E_{BV} \approx 7 \times 10^{37}$ \\
UU CrB & Photometry & 8.63 & $\Delta I = 0.30 mag$ & $> \sim 57$ min &
$E_{opt} = 7 \times 10^{35}$ \\
\hline
\end{tabular}
\end{center}
\end{table}
 
\begin{table}
\caption{{\bf Are the stars like our Sun?}}
\begin{center}
\begin{tabular}{lllllll}
\hline
Star & Spectrum & Li EW & $v \cdot sin(i)$ & S & $L_{X}$ & Companion \\
     &          & (m$\AA$) & ($km \cdot s^{-1}$) & & ($erg \cdot
s^{-1}$) & \\
\hline
Gmb 1830 & G8 V & 4.0 & 1.3 & 0.188 & $< 4 \times 10^{27}$ & Single \\
$\kappa$ Cet & G5 V & 38 & 8 & 0.366 & $8.2 \times 10^{28}$ & V = 9.3,
a = 269 ''\\
MT Tau & G5 V & $\cdots$ & $\cdots$ & $\cdots$ & $< 2 \times 10^{32}$ &
Single \\
$\pi^{1}$ UMa & G1.5 Vb & 96 & 9.7 & 0.367 & $9.3 \times 10^{28}$ & Single
\\
S For & G1 V & 34 & 7 & $\cdots$ & $< 2 \times 10^{29}$ & V = 9.3, a =
0.3'' \\
BD $+10^{\circ}2783$ & G0 V & $< 23$ & 4 & $\cdots$ & $< 9 \times 10^{29}$
& Single \\
o Aql & F8 V & 63 & 3.9 & 0.148 & $< 2 \times 10^{28}$ & V = 13.4, a =
22''\\
5 Ser & F8 IV-V & $< 3$ & 2 & 0.140 & $< 3 \times 10^{28}$ & V = 9.7,
a = 11''\\
UU CrB & F8 V & $< 12$ & 6 & $\cdots$ & $< 3 \times 10^{29}$ & Single \\
\hline
\end{tabular}
\end{center}
\end{table}

\end{document}